\documentclass[prd,twocolumn,showpacs,amsmath,amssymb]{revtex4}
\usepackage{graphicx}
\begin{document}
\title{Friedmann cosmology with bulk viscosity: a concrete model for dark energy}
\author{Xin-He Meng}
\email{xhm@nankai.edu.cn}
\author{Xu Dou}
\email{dowxdou@gmail.com}
\affiliation{Department of physics,
Nankai University, Tianjin 300071, China}
\date{\today}




\begin{abstract}
The universe content is considered as a non-perfect fluid with bulk
viscosity and can be described by a  general equation of state
(endowed some deviation from the conventionally assumed cosmic
perfect fluid model). An explicitly bulk viscosity dark energy model
is proposed to confront consistently with the current observational
data sets by statistical analysis and is shown consistent with (not
deviated away much from) the concordant $\Lambda$ Cold Dark Matter
(CDM) model by comparing the decelerating parameter. Also we compare
our relatively simple viscosity dark energy model with a more
complicated one by contrast with the concordant $\Lambda$CDM model
and find our model improves for the viscosity dark energy model
building. Finally we discuss the perspectives of dark energy probes
for the coming years with observations.
\end{abstract}




\maketitle

\section{Introduction}
The cosmological observations indicate that the expansion of our
universe accelerates \cite{bah9}. Recently£¬ lots of research work
on its possible mechanism, such as extended gravity \cite{mw1},
decaying $\Lambda$ Cold Dark Matter (CDM) model\cite{pm},
modifying equation of state (hereafter EOS)or by introducing kinds
of the so called dark energy models are to explain the cosmic
acceleration expansion observed. To consider causality and the
hydrodynamical instability, an interesting barotropic dark energy
model is proposed\cite{ls} that includes a linear EOS of with a
general form, $p=\alpha(\rho-p_0)$ \cite{bab05}, which
incorporated into cosmological model can describe the
hydrodynamically stable dark energy behaviors.

The astrophysical observations also indicate that the universe
media is not a perfect fluid \cite{jaf05} and the viscosity is
concerned in the evolution of the universe
\cite{bre05a,bre05b,cat05}. On the other hand, in the standard
cosmological model, if the EOS parameter $\omega$ is less than
$-1$, the universe shows the future finite singularity called Big
Rip \cite{cal03,noj05}. Several ideas are proposed to prevent the
big rip singularity as thought it un-physical, like by introducing
quantum effects terms in the action \cite{od}, or by including
universe viscosity media for the Universe evolution\cite{rm}. So
an interesting question naturally arises: what kind of role the
cosmic viscosity element can play for helping the above two
facets, dark energy and cosmic dark energy model singularity?
Considering some deviation from the ideal fluid model is also very
helpful to nowadays cosmology probes advancement.

In Refs.~\cite{zim96,col96,chi97,mak98,zim00,pri00,maa96}, the
bulk viscosity in cosmology has been studied in various aspects.
Dissipative processes are thought to be present in any realistic
theory of the evolution of the universe. In the early universe,
the thermodynamics is far from equilibrium, the viscosity should
be concerned in the studies of the early stage for cosmological
evolution. But even in the later cosmic evolution  stage, for
example, the temperature for the intergalactic medium (IGM),
baryonic gas, generally is about 10000K to 1000000K and the
complicated IGM is rather non-trivial. The sound speed $c_s$ in
the baryonic gas is only a few km$s^{-1}$ to a few tens km$s^{-1}$
and the Jeans length $\lambda$ yields a term as an effective
viscosity $c_s \lambda $. On the other hand , the bulk velocity of
the baryonic gas is of the order of hundreds km$s^{-1}$ \cite{pb}.
So it is helpful to consider the viscosity element in the later
cosmic evolution. It is well known that in the framework of
Friedmann Robertson Walker (FRW) metric, the shear viscosity has
no contribution in the energy momentum tensor, and the bulk
viscosity behaves like an effective pressure. At the late times,
since we do not know the nature of the universe content (dark
matter and dark energy components) very clearly, concerning the
bulk viscosity is reasonable and practical. Moreover, the cosmic
viscosity here can also be regarded as an effective quantity as
caused by complicated astrophysics mechanisms and may play a role
as a dark energy candidate \cite{zim01,ren05,hm} or a possible
unification scheme for the two mysterious dark components (dark
matter and dark energy)\cite{hm,const, con} as they may be facets
of the same problem, and then evidence for dark matter is also
evidence for dark energy.

The EOS $p=(\gamma-1)\rho$ and the bulk viscosity
$\zeta=\alpha\rho^s$ is studied in the full causal theory of bulk
viscosity, and the case $s=1/2$ has possessed exact solutions as
shown in \cite{chi97,mak98}. However, both the pressure and the
bulk viscosity coefficient may have constant components. We argue
that the non-causal approximation is reasonable in the late times
of the universe evolution. In our previous papers, we have shown
that the Friedmann equations can be solved with both a more
general EOS and bulk viscosity detailed as follows.(Note that the
EOS with a constant pressure gets degenerate with $\Lambda$CDM
model.)
\begin{equation}
p=(\gamma-1)\rho+p_0,
\end{equation}
where $p_0$  and $\gamma$ are two parameters. The bulk viscosity
is expressed as
\begin{equation}
\zeta=\zeta_0+\zeta_1\frac{\dot{a}}{a}.
\end{equation}
where $\zeta_0$ and $\zeta_1$ are two constants conventionally, and
the overhead dot stands for derivative with respect to time. The
motivation of considering this bulk viscosity is that by fluid
mechanics we know the transport/viscosity phenomenon is involved
with the "velocity" $\dot{a}$, which is related to the scalar
expansion $\theta=3\dot{a}/a$. Both $\zeta=\zeta_0$ (constant) and
$\zeta\propto\theta$ are considered in the previous papers
\cite{bre02,bre05a}, so a linear combination of the two are more
general. The $\omega=p/\rho$ is constrained as $-1.38<\omega<-0.82$
by present observation data, so the inequality in our previous case
should be
\begin{equation}
-1.38<\gamma-1+\frac{p_0}{\rho}<-0.82.
\end{equation}
The parameter $p_0$ can be positive (attractive force) or negative
(repulsive force), and conventionally $\zeta_0$ and $\zeta_1$ are
regarded as positive. To choose the parameters properly, it can
prevent the Big Rip problem or some kind of singularity for the
cosmology model, like in the phantom energy phase, as demonstrated
before. Additionally, the sound speed in this model can also keep
the casuality condition.

This present paper is a continuous work following our previous
efforts, which is organized as follows. In Sec. II we present a
relatively simple cosmology model with extremely non-relativistic
dark matter and viscosity dark energy by an explicit form. With
these we give out the exact solution and discuss the acceleration
phase in this model. In the last section (Sec.III ) we discuss and
summarize our conclusions.

\section{Model and calculations}
We consider the Friedmann-Roberson-Walker metric in the flat space
geometry ($k$=0) as the case favored by WMAP satellite mission on
cosmic background radiation (CMB) data
\begin{equation}
ds^2=-dt^2+a(t)^2(dr^2+r^2d\Omega^2),
\end{equation}
and assume that the cosmic fluid possesses a bulk viscosity. The
energy-momentum tensor can be written as
\begin{equation}
T_{\mu\nu}=\rho U_\mu U_\nu+(p+\Pi)H_{\mu\nu},
\end{equation}
where in the co-moving coordinates $U^\mu=(1,0)$, and
$H_{\mu\nu}=g_{\mu\nu}+U_\mu U_\nu$ \cite{bre02}. By defining the
effective pressure as $\tilde{p}=p+\Pi$ and from the Einstein
equation $R_{\mu\nu}-\frac{1}{2}g_{\mu\nu}R=8\pi GT_{\mu\nu}$, we
obtain the Friedmann equations
\begin{subequations}
\begin{eqnarray}
\frac{\dot{a}^2}{a^2} &=& \frac{8\pi G}{3}\rho\label{eq1},\\
\frac{\ddot{a}}{a} &=& -\frac{4\pi
G}{3}(\rho+3\tilde{p})\label{eq2}.
\end{eqnarray}
\end{subequations}
The covariant conservation equation for energy $T^{0\nu}_{;\nu}$,
yields
\begin{equation}
\dot{\rho}+(\rho+\tilde{p})\theta=0,
\end{equation}
where the expansion parameter $\theta=U^\mu_{;\mu}=3\dot{a}/a$.

Then the Friedmann equations give:
\begin{displaymath}
8\pi G\rho=3\left(\frac{\dot{a}}{a}\right)^2=3H^2
\end{displaymath}
If we assume that the cosmic fluid possesses a bulk viscosity as
shown explicitly in the pressure as $\Pi=-\xi\theta$ (we take
another Greek character to represent the bulk viscosity which is
different from the before papers), the energy-momentum tensor
could be written fully as :
\begin{displaymath}
T_{\mu\nu}=\rho U_{\mu}U_{\nu}+\left(p-\xi\theta \right)H_{\mu\nu}
\end{displaymath}
where $U_{\mu}$ denotes 4 - velocity, as before defined $\theta
=3H$, and
 ${\xi}$ is the bulk viscosity with its explicit form to be present below.

The cold dark matter has been assumed to be extremely
non-relativistic so that we can take its pressure $p=0$, and we
now suppose that the effect of dark energy on cosmos evolution is
included in the viscous term $-\xi\theta$ which has the dimension
of pressure. In this present work we treat it
as the effective pressure of dark energy. \\
Then the covariant conservation of $T_{\mu\nu}$ (Eq.(7)) yields:
\begin{displaymath}
\dot\rho+\left(\rho-\xi\theta\right)\theta=0
\end{displaymath}
so,
\begin{equation}\label{eq1}
\dot\rho+\rho\theta=\xi{\theta}^{2}
\end{equation}
Here, it will be convenient to define a dimensionless parameter
below:
\begin{displaymath}
h^{2}=\frac{H^{2}}{{H_0}^{2}}=\frac{\rho}{\rho_{cr}}
\end{displaymath}
where $\rho_{cr}$ is the critical density. Using this definition, we
could rewrite equation (\ref{eq1}) as:
\begin{equation}\label{eq2}
\frac{d{h^{2}}/dt}{H_{0}}+3h^{3}=9\lambda h^{2}
\end{equation}
where $\lambda=\xi H_{0}/\rho_{cr} $ is the bulk viscosity.

For comparison with the observational data, we should derive $h$
from Eq.(\ref{eq2}) by writing it as the equation of red-shift $z$
as:
\begin{equation}\label{eq3}
-2(1+z)\frac{dh}{dz}+3h=9\lambda
\end{equation}
In the last part of \cite{const}, condition with constant
viscosity has been considered. It proves that it has a good
fitting with SNe Ia data sets. Recently, a complicated viscosity
form dependent on $h$ has been proposed in ref.\cite{c}. To
overcome old cosmological constant problem we believe that a
variable viscosity will give a better result. Here we propose an
explicit red-shift dependent bulk viscosity, which gets a constant
limit (an effective cosmological constant) today as $z=0$. And we
will show that this relatively simple form is better than the
complex one in ref.\cite{c} when compared with $\Lambda$CDM model.
The explicit red-shift dependent viscosity is
\begin{equation}
9\lambda=\lambda_{0}+\lambda_{1}(1+z)^n
\end{equation}
where $n$ is an arbitrary integer, $\lambda_{0}$, and
$\lambda_{1}$ are two arbitrary constants, which could all be best
fitted from the observational data sets.

With this new bulk viscosity, we could write Eq. (\ref{eq3}) as:
\begin{equation}
-2(1+z)\frac{dh}{dz}+3h=\lambda_{0}+\lambda_{1}(1+z)^n
\end{equation}
which is a first-order differential equation of $h$ and the exact
solution of $h$ from this equation is:
\begin{equation}
h=\lambda_{2}(1+z)^{1.5}-\frac{\lambda_{1}}{2n-3}(1+z)^{n}+\frac{\lambda_{0}}{3}
\end{equation}
where $\lambda_{2}$ is an integration constant. The obtained
relation above could be regarded that the cosmic expansion rate is
from a combined result of the viscosity term: the matter component
effect plus the effective
 dark energy (including the cosmological constant like term $\frac{\lambda_{0}}{3}$)
contributions.  Because of the consistent requirement $h=1$ for
the spatial flat universe at present, from the above we have
\begin{displaymath}
\frac{\lambda_{0}}{3}=1-\lambda_{2}+\frac{\lambda_{1}}{2n-3}
\end{displaymath}
 With the analysis of dimension of term
$\lambda_{2}$ and the fact that matter density of our cosmos
$\rho_{m}\propto a(t)^{-3}$, where $a(t)$ is the scalar scale
factor we will show that the value of $\lambda_{2}$ from data
fitting below does consist with the result of matter component
given by WMAP5 data sets\cite{w}. To make the fitting result
compared with $\Lambda$CDM model, we make an assumption that
$\lambda_{2}=\sqrt{\Omega_{m0}}$, where $\Omega_{m0}\equiv8\pi
G\rho_{m0}/(3{H_{0}}^2)$. So in our viscosity dark energy model,
the first term as labelled by $\lambda_{2}$ could be looked as
matter contribution and the other terms of $h(z)$ is due to the
viscosity dark energy, which with limit case ($z=0$ today) as the
concordant $\Lambda$CDM model. The consistent requirement relation
also gets clear meaning, that is, the relationship between the
matter and dark energy (including the cosmological constant
contribution) components.

The observations of the SNe Ia have provided the first direct
evidence of the accelerating expansion for our current universe.
so any model attempting to explain the acceleration mechanism
should be consistent with the SNe Ia data implying results, as a
basic requirement. As we know the observations of supernovas
measure essentially the apparent magnitude $m$, which is related
to the luminosity distance $d_{L}$ by
\begin{equation}
m=M+5\log_{10}D_{L}\left(z\right)
\end{equation}
where the distance 
$D_{L}\left(z\right)\equiv\left(H_{0}\right)d_{L}\left(z\right)$
is the dimensionless luminosity and
\begin{equation}
d_{L}=(1+z)d_{M}(z),
\end{equation}
where $d_{M}$ is the co-moving distance given by
\begin{equation}
d_{M}=\int_{0}^{z}\frac{1}{H(z^{'})}dz^{'}
\end{equation}
Also,
\begin{equation}
\mathcal{M}=M+5\log_{10}\left(\frac{1/H_{0}}{1Mpc}\right)+25,
\end{equation}
where $M$ is the absolute magnitude which is believed to be constant
for all supernovaes of type Ia. In this paper, we use the 307 Union
SNe Ia data sets compiled in reference\cite{un} . The data points in
these samples are given in terms of the distance modulus

\begin {equation}
\mu_{obs}\equiv m(z)-M_{obs}(z)
\end{equation}
We employ it for doing the standard statistic analysis. So the
$\chi^{2}$ is calculated from
\begin{equation}
\chi^{2}=\sum_{i=1}^{n}\bigg[\frac{\mu_{obs}(z_{i})-M^{'}-5\log_{10}D_{Lth}(z_{i};c_{\alpha})}{\sigma_{obs}(z_{i})}\bigg]^{2}
\end{equation}
where $M^{'}=\mathcal{M}-M_{obs}$ is a free parameter and
$D_{Lth}(z_{i};c_{\alpha})$ is the theoretical prediction for the
dimensionless luminosity distance of a supernovae at a particular
distance, for a given model with parameter $c_{\alpha}$.

On the other hand, the shift parameter $\mathcal{R}$ and the
distance parameter $\mathcal{A}$ are considered to give
contributions to data fitting. The shift parameter $\mathcal{R}$
is defined in refs.\cite{A} and \cite{ap} as
\begin{equation}
\mathcal{R}\equiv\sqrt{\Omega_{m}}\int_{0}^{z_{*}}\:\frac{d
z^{'}}{h(z^{'})}
\end{equation}
and WMAP5 results \cite{w} have updated the redshift of
recombination to be $z_{*}=1090$. Its detail meaning can be found
in reference \cite{a}. The distance parameter $\mathcal{A}$ is
defined as
\begin{equation}
\mathcal{A}\equiv\sqrt{\Omega_{m}}\:h(z_{b})^{-\frac{1}{3}}\big(\frac{1}{z_{b}}\int^{z_{b}}_{0}\:\frac{d
z^{'}}{h(z^{'})})^{\frac{2}{3}}
\end{equation}
where $z_{b}=0.35$.

Considering $\mathcal{R}$ and $\mathcal{A}$, we use the total
$\chi^{2}$ to make the standard statistic analysis, and data
fitting:
\begin{equation}
\chi^{2}_{total}=\chi^{2}+\left(\frac{\mathcal{R}-\mathcal{R}_{obs}}{\sigma_{\mathcal{R}}}\right)^{2}+\left(\frac{\mathcal{A}-\mathcal{A}_{obs}}{\sigma_{\mathcal{A}}}\right)^{2}
\end{equation}

\begin{figure}
\includegraphics[scale=0.6]{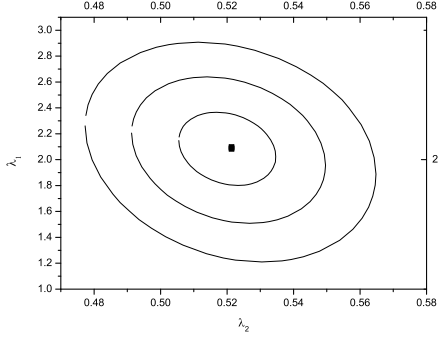}
\caption{The $1\sigma$, $2\sigma$, $3\sigma$ C.L. contours of
 $\lambda_{1}$ and $\lambda_{2}$ in the best fitting condition $n=-1$. The black
 dot corresponds to
the best fitting values.}
\end{figure}

The result of data fitting is listed in TABLE 1.

\begin{table}
\centering \caption{Fitting results for model parameters}
\begin{tabular}{ccccc}\\
\hline\hline
n &$\lambda_{0}$ &$\lambda_{1}$ &$\lambda_{2}$ &$\chi^{2}$\\
\hline
-1 &    0.18027 &    2.0923 &    0.52145 &    312.8\\
\hline -0.8 &    -0.04672 &    2.2784 &    0.52027 &    313.0747\\
\hline -2 &    0.63824 &    1.8331 &    0.52538 &    312.2193\\
\hline 1 &    1.0464 &    -0.15136 &    0.50016 &    323.3296\\
\hline

\end{tabular}
\end{table}

In Fig. 2, we plot the deceleration parameter $q$ relation with
the redshift as model comparisons:
\begin{equation}
q=\frac{1+z}{H}\:\frac{dH}{dz}-1
\end{equation}
At the same time, we plot the deceleration parameter relations of
$q(z)$ in
$\Lambda$CDM model and the more complex bulk viscosity in ref.\cite{c} for contrast in Fig. 2, too.\\

\begin{figure}
\includegraphics[scale=0.8]{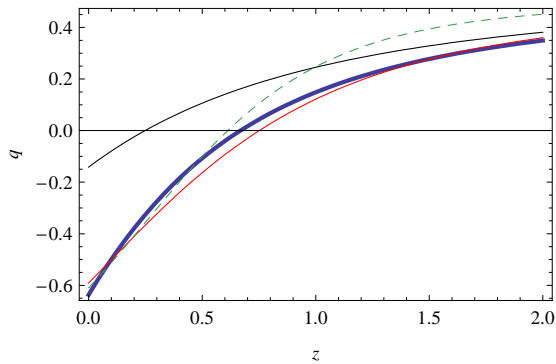}
\caption{$q-z$ relation. Red line (the lowest one) represents
$\Lambda$CDM, blue thick line and black line are our models with
$n=-1$ and $n=1$ respectively. By including the more complex
viscosity form in \cite{c}, we get another result(dashed line) for
illustration.}
\end{figure}
To demonstrate clearly how and when the viscosity dark energy
components catch up of the matter contribution, and overpass it we
also plot the two components evolution in Fig.3
\begin{figure}
\includegraphics[scale=0.8]{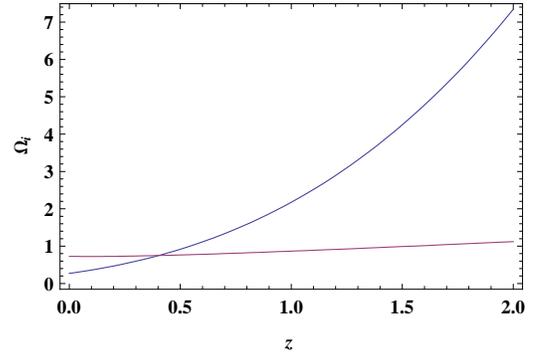}
\caption{The bulk viscosity dark energy part (the slow changing
almost horizontal line) evolutions vs matter component}
\end{figure}
 With our best fitting result of the $n=-1$ case, we could calculate the  EOS of viscosity dark
energy. The pressure of dark energy is $-\xi\theta$, so state
parameter:
\begin{equation}
\omega=-\frac{\xi\theta}{\rho_{X}}
\end{equation}
where $\rho_{X}=\rho(1-\Omega_{m})$ is the density of dark
energy.\\
Then we have,
\begin{equation}
\omega=-\frac{3H\xi}{\rho(1-\Omega_{m})}
\end{equation}
With the relation of $\xi$ and $\lambda$ defined above, we could get
$\omega$ today:
\begin{equation}
\omega_{0}=-\frac{\lambda_{0}+\lambda_{1}}{3(1-\Omega_{m0})}
\end{equation}
By the best fitting data we give an approximated value of EOS at
present($h=1$) $\omega\simeq-1.04$. There is a little deviation
from $\Lambda$CDM model by a value of -0.04.

Recently, ref.\cite{dia} has proposed $Om$ diagnose method to
differentiate a new model from the $\Lambda$CDM model with the
constant equation of state parameter exactly as -1. The diagnostic
parameter $Om$ is defined as:
\begin{equation}
Om=\frac{h^{2}-1}{x^3-1}
\end{equation}

where $x=1+z$. The dimensionless expansion parameter $h(z)$ for a
dark energy model with constant equation of state can be written
clearly as:
\begin{equation}
h(x)^{2}=\Omega_{m0}x^{3}+(1-\Omega_{m0})x^{3(1+\omega)}
\end{equation}
So, for the $\Lambda$CDM model with its EOS parameter $\omega=-1$,
$Om=\Omega_{m0}$. This result also gives us a $null-test$ of
cosmology constant. For model we proposed(here we use the best
fitting results with $n=-1$, which gives a better $q(z)$ evolution
compared with the $\Lambda$CDM model), its $h^{2}$ could be written
as:
\begin{equation}
h^2=\big(\lambda_{2}x^{1.5}+\frac{\lambda_{1}}{5}x^{-1}+\frac{\lambda_{0}}{3})^2
\end{equation}
so the $Om$ evolution for this viscosity dark energy model is
shown in Fig.4 to compare with the $\Lambda$CDM model which is a
horizontal line in the plot.

\begin{figure}
\includegraphics[scale=0.8]{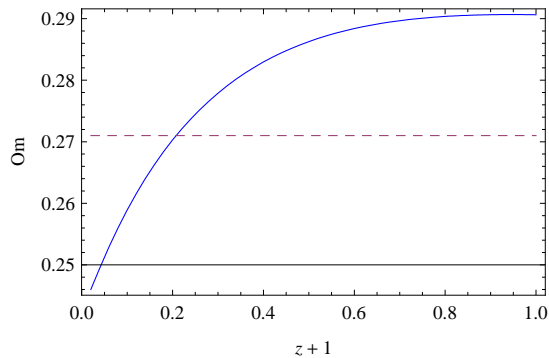}
\caption{$Om-z$ relation as clearly shown the viscosity dark
energy model deviations from $\Lambda$CDM model. The dashed line
corresponds $\Lambda$CDM model while the thick curve corresponds
to our viscosity dark energy model contribution.}
\end{figure}

With this diagnostic method, we could also see that this model has
different properties from $\Lambda$CDM model with the exact
constant EOS parameter $\omega=-1$.

With the above discussions we can see our relative simple
viscosity model fits the concordant $\Lambda$CDM model better. It
is due to the simpler viscosity form. As a matter of fact so far
we still have not obtained a systematic way to express the cosmic
viscosity effects. In view of the astrophysical reality any
progresses in this field is very helpful. To show the similarities
and differences among the concordant $\Lambda$CDM model, our model
and the complicated one \cite{c} we present a much clear figure
below with the hope there will be much relative works to appear
soon. Hope the study for viscosity dark energy can be stimulated
further.

\begin{figure}
\includegraphics[scale=0.8]{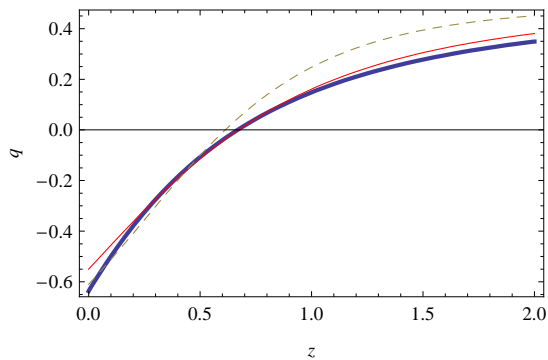}
\caption{$q(z)-z$ relations for three different models. The nearby
two lines are ours with the explicit viscosity form and the
concordant $\Lambda$CDM model($\Omega=0.3$). The dashed line
corresponds the complicated one, which can degenerate the previous
two models for certain redshift values}
\end{figure}

\section{Summary and discussions}
In this letter we present an explicit viscosity form to mimic dark
energy behaviors and confront it with current observational data
sets. Though the viscosity dark energy form is simple, the results
are better. The possible Universe evolution fate, like possible
future singularity types from this viscosity model can be also
discussed and we will work it out elsewhere. We emphasize that
perfect fluid is just a limit case of a general viscosity media that
is more practical in the astrophysics sense.

Discovery of dark energy is about ten years old, but its nature and
origin have been still puzzling. Fundamental as it has promised to
physics foundations there are several possibilities to develop with
great expectations. One of these is the unification scenario of dark
matter and dark energy, that is, they are two facets of one secret.
If we finally discover the dark matter either by accelerators such
as LHC or future ILC or by satellite missions, like PAMELA and GLAST
now on the running, we may infer the existence of dark energy
therefore. In this aspect viscosity dark energy has already shown
its unification efforts, to write uniformly all the possible cosmic
components in a formula with distinct scaling law with respect to
scalar factor\cite{const, con,hm}. It is worthy further studying.

\section*{Aknowledgement}
 We thank Profs. I. Brevik , S.D.
Odintsov and Lewis H. Ryder for lots of interesting discussions
during the project. This work is partly supported by NSF of China
under Grant No.10675062


\begin{thebibliography}{99}

\bibitem{bah9} T. Totani, Y. Yoshii, and K. Sato, Astrophys. J.
\textbf{483}, L75 (1997); S. Perlmutter \textit{et al.}, Nature
\textbf{391}, 51 (1998); A.G. Riess \textit{et al.}, Astron. J.
\textbf{116}, 1009 (1998); N. Bahcall, J.P. Ostriker, S.
Perlmutter, and P.J. Steinhardt, Science \textbf{284}, 1481
(1999).
\bibitem{mw1} X.H.Meng and P.Wang, Class. Quant.Grav.20,4949(2003);
ibid, 21, 951(2004); ibid, 21, 2029(2004); ibid,22, 23(2005);
ibid, Gen.Rel.Gra.(2004)36, 1947; ibid,  Phys. Lett. B584,
1(2004); E.Flanagan, Class.Quant.Grav.21, 417 (2003); S. Nojiri
and S. Odintsov, Phys.Lett.B576,5 (2003); ibid, Phys. Rev. D68,
123512 (2003); D. Vollick, ibid, D68, 063510(2003);G.Ellis,
arXiv:0811.3529[astro-ph] and references therein; to an incomplete
list,
\bibitem{pm}P.Wang and X.H.Meng,Class. Quant.Grav.22(2005)283; U Mukhopadhyay, S
Ray and X.H.Meng ,
 Gravitation and  Cosmology 13 (2007) 142;  X.H. Meng,et.al, in
 preparations
\bibitem{ls}E.Linder and R.Scherrer, arXiv:0811.2797[astro-ph]
\bibitem{bab05} E. Babichev, V. Dokuchaev, and Y. Eroshenko, Class.
Quantum Grav. \textbf{22}, 143 (2005) .
\bibitem{jaf05} T.R. Jaffe, A.J. Banday, H.K. Eriksen, K.M.
G\'{o}rski, and F.K. Hansen, astro-ph/0503213.
\bibitem{bre05a} I. Brevik and O. Gorbunova, gr-qc/0504001.
\bibitem{bre05b} I. Brevik, O. Gorbunova, and Y. A. Shaido, gr-qc/0508038.
\bibitem{cat05} M. Cataldo, N. Cruz, and S. Lepe, Phys. Lett. B
\textbf{619}, 5 (2005).
\bibitem{cal03} R.R. Caldwell, M. Kamionkowski, and N.N. Weinberg,
Phys. Rev. Lett. \textbf{91}, 071301 (2003).
\bibitem{noj05} S. Nojiri, S.D. Odintsov, and S. Tsujikawa, Phys.
Rev. D \textbf{71}, 063004 (2005).
\bibitem{od}S. Nojiri, and S.D. Odintsov, Phys.Lett.B595,1(2004);
E.Elizalde,S. Nojiri, and S.D. Odintsov, Phys.Rev.D70,
0343539(2004); for example.
\bibitem{rm}X.H.Meng,J.Ren and M.Hu, astro-ph/0509250, Comm.Theor.Phys.47(2007)379 ,
\bibitem{zim96} W. Zimdahl, Phys. Rev. D \textbf{53}, 5483 (1996).
\bibitem{col96} A.A. Coley, R.J. van den Hoogen, and R. Maartens,
Phys. Rev. D \textbf{54}, 1393 (1996).
\bibitem{chi97} L.P. Chimento, A.S. Jakubi, V. M{\'e}ndez, and
R. Maartens, Class. Quantum Grav. \textbf{14}, 3363 (1997).
\bibitem{mak98} M.D. Mak and T. Harko, J. Math. Phys. \textbf{39},
5458 (1998).
\bibitem{zim00} W. Zimdahl, Phys. Rev. D \textbf{61}, 083511 (2000).
\bibitem{pri00} A.D. Prisco, L. Herrera, and J. Ib{\'a}{\~n}ez, Phys. Rev. D
\textbf{63}, 023501 (2000); T. Padmannabhan and S.Chitre,
Phys.Lett.A120, 433(1987).
\bibitem{maa96} R. Maartens, astro-ph/9609119.
\bibitem{pb}P.Peebles, Principles of Physical Cosmology, Princeton
University Press (1993); E.Kolb and M.Turner, The Early Universe,
Addison-Wesley (1990): J.Peacock, Cosmological Physics, Cambridge
University Press (1999); A.Liddle and D.Lyth, Cosmological
Inflation and Large-Scale Structure, Cambridge University Press
(2000). S.Weinberg, Gravitation and Cosmology, John Wiley and Sons
(1972). G.Borner, The Early Universe, Facts and Fiction,
Springer-Verlag (1988).
\bibitem{zim01} W. Zimdahl, D. J. Schwarz, A. B. Balakin, D. Pavon,
Phys. Rev. D \textbf{64}, 063501 (2001), astro-ph/0009353.
\bibitem{ren05} J. Ren and X.H. Meng, astro-ph/0511163, Phys. Lett.
B.636(2006)5
\bibitem{hm} M.G.Hu and X.H.Meng, astro-ph/0511615; Phys. Lett.
B.635(2006)186.
\bibitem{bre02} I. Brevik, Phys. Rev. D \textbf{65}, 127302 (2002).
\bibitem{un} M. Kowalski \textsl{et al}, arXiv:0804.4142[astro-ph].
\bibitem{dia} V. Sahni, A. Shafieloo and A. A. Starobinsky Phys. Rev. D 78, 103502 (2008) arXiv:0807.3548[astro-ph].
\bibitem{w} E. Komatsu \textsl{et al}, arXiv:0803.0547[astro-ph].
\bibitem{A} J. R. Bond, G. Efstathiou and M. Tegmark, Mon. Not. Roy.
Astron. Soc. 291, L33 (1997).
\bibitem{ap} Y. Wang and P. Mukherjee, Astrophys. J. 650, 1 (2006)
\bibitem{a} D.J. Eisenstein \textsl{et al.}, Astrophys. J. 633, 560 (2005).
\bibitem{c} V. Folomeev and V. Gurovich, Phys. Lett. B 661(2008) 75-77
\bibitem{const} J. Ren and X.H. Meng, Phys. Lett. B 633 (2006) 1.
\bibitem{con}J.Ren and X.H. Meng, 
astro-ph/0605694;Int.J.Mod.Phys.D16 (2007) 1341,

J.Ren, X.H. Meng and L Zhao, 
Phys Rev.D76 (2007)043521
\end{thebibliography}
\end{document}